\documentclass[aps,prd,floatfix,nofootinbib,showpacs,twocolumn,superscriptaddress]{revtex4-1}
\usepackage{amsmath}
\usepackage{amssymb}
\usepackage{bbold}
\usepackage{bm}
\usepackage{color}
\usepackage{graphicx}
\usepackage{mathrsfs}
\usepackage{soul}

\usepackage[mathscr,scaled=1.15]{urwchancal}
\DeclareFontFamily{OT1}{pzc}{}
\DeclareFontShape{OT1}{pzc}{m}{it}%
{<-> s * [1.15] pzcmi7t}{}
\DeclareMathAlphabet{\mathpzc}{OT1}{pzc}{m}{it}

\definecolor{purple}{rgb}{0.5,0,0.5}
\definecolor{blue}{rgb}{0.0,0,0.9}
\usepackage[hypertex]{hyperref}
\hypersetup{colorlinks=true, linkcolor=purple, citecolor= purple, urlcolor=blue}

\newcommand{\tr}{\mathrm{tr}_\mathrm{CD}\!\ }

\begin{document}

\title{Couplings between the $\rho$ and $D$- and $D^\ast$-mesons}

\author{Bruno~El-Bennich}
\affiliation{Laboratorio de F\'isica Te\'orica e Computa\c{c}\~ao, Universidade Cruzeiro do Sul, 01506-000, S\~ao Paulo, Brazil}

\author{M.~Ali Paracha}

\affiliation{Department of Physics, School of Natural Sciences, National University of Science and Technology, Islamabad, Pakistan}

\author{Craig~D.~Roberts}
\affiliation{Physics Division, Argonne National Laboratory, Argonne, Illinois 60439, USA}

\author{Eduardo Rojas}

\affiliation{Instituto de F\'isica, Universidad de Antioquia, Calle 70, no.~52-21, Medell\'in, Colombia}

\date{21 November 2016}

\begin{abstract}
We compute couplings between the $\rho$-meson and $D$- and $D^\ast$-mesons -- $D^{(\ast)}\!\rho D^{(\ast)}$ -- that are relevant to phenomenological meson-exchange models used to analyse nucleon--$D$-meson scattering and explore the possibility of exotic charmed nuclei.  Our framework is built from elements constrained by Dyson-Schwinger equation studies in QCD, and therefore expresses a simultaneous description of light- and heavy-quarks and the states they constitute. We find that all interactions, including the three independent $D^{\ast} \rho \,D^{\ast}$  couplings, differ markedly amongst themselves in strength and also in range, as measured by their evolution with $\rho$-meson virtuality.  As a consequence, it appears that one should be cautious in using a single coupling strength or parametrization for the study of interactions between $D^{(\ast)}$-mesons and matter.
%
\pacs{
14.40.Lb,   
13.25.Ft,   
11.15.Tk,   
12.39.Ki    
}
\end{abstract}

\maketitle

\section{Introduction}
\label{intro}
Development of a detailed understanding of charmonium production and decay in heavy-ion collisions is an important but difficult task.  The charmonium production rate is correlated with the nature of the medium produced in the heavy-ion collision, which might be a quark gluon plasma; and many of the charmonium decays involve production of mesons with exposed charm, which themselves interact both with lighter mesons and the medium on their way to producing the observed final states.  Detailed knowledge of such hadronic effects and interactions is therefore required in order to reach reliable conclusions about a new state of matter from the final composition and distribution of decay products.

Information about charm-meson final-state interactions can be inferred from charmonium decays.  An important example is the ``exotic'' heavy charmed state $X(3872)$, first observed in $B$ decays by Belle \cite{Choi:2003ue}, confirmed by BaBar \cite{Aubert:2004ns}, and found in $\bar pp$ collisions by the CDF~II \cite{Acosta:2003zx} and D$\O$ \cite{Abazov:2004kp} Collaborations.  The $X(3872)$ appears to be a $J^{PC}=1^{++}$ state and, within errors, its mass \cite{Agashe:2014kda}, $M_X =3871.69 \pm 0.12\,$MeV$/c^2$, coincides with the $\bar D^0 D^{*0}$ threshold. Its composition is controversial and numerous theoretical descriptions have been proposed, \emph{e.g}.\ a mixture of pure charmonium and a molecular bound state \cite{Matheus:2009vq}, a purely molecular bound state \cite{Swanson:2004pp, Liu:2006df, Fleming:2007rp, Gamermann:2009fv, Dias:2011mi, Torres:2014fxa, Tomaradze:2015cza}, and a tetraquark state \cite{Dubnicka:2010kz, Narison:2010pd, Chen:2010ze, Cui:2011be}.  Molecular bound states are plausible since the mass difference between the $X(3872)$ and the $\bar D^0 D^{*0}$ system is extremely small, as discussed in Ref.\,\cite{Tomaradze:2015cza}; and although the binding mechanisms differ, these states might bear similarities to conjectured $\bar pp$ bound states in $J/\psi$ decays \cite{ElBennich:2008vk, Dedonder:2009bk}.

Irrespective of its internal structure, the observed decays $X(3872)\to J/\psi\,\pi^+\pi^-$ \cite{Choi:2003ue, Aubert:2004ns, Acosta:2003zx, Abazov:2004kp}, $X(3872)\to J/\psi\,\pi^+\pi^-\pi^0$ , $X(3872)\to J/\psi\,\gamma $ \cite{Abe:2005ix}, and $X(3872)\to D^0 \bar D^0 \pi^0$ \cite{Gokhroo:2006bt} involve final-state interactions, which can be studied using phenomenological Lagrangians.  In such an approach, the $J/\psi\, \rho$ state in $X(3872)\to J/\psi\, \rho \to J/\psi\, \pi^+\pi^-$ is preceded by a $D$-meson loop that couples to the $J/\psi$ and $\rho$ \cite{Liu:2006df} and the reaction $D^{(*)}D^{(*)} \to \pi X(3872)$ involves an intermediate triangle diagram with pseudoscalar and vector mesons \cite{Torres:2014fxa}.  The model Lagrangians are expressed in terms of couplings, \emph{e.g}.\ $g_{D^* \pi D}$, $g_{D\rho D}$, $g_{D^\ast \!\rho D}$ and $g_{D^\ast\!\rho D^\ast}$, which are \emph{a priori} unknown.

Heavy-quark symmetry implies that all such Lagrangian couplings are degenerate at leading order \cite{Casalbuoni:1996pg, Manohar:2000dt}; but significant $\Lambda_\mathrm{QCD}/m_c$ corrections spoil this prediction.  In the limit of exact chiral, heavy-flavor and spin symmetries, a heavy-meson chiral effective Lagrangian \cite{Casalbuoni:1996pg} describes the strong interactions between any two heavy mesons and a pseudoscalar Nambu-Goldstone boson, where the effective coupling, $\hat g$, plays a basic role.  At leading order in the effective theory, this coupling can be related to the hadronic $\langle \pi D | D^*\rangle$ decay amplitude:
\begin{equation}
     \hat g = \frac{f_\pi}{2 M_D} \, g_{D^\ast \pi D} \ ,
\label{effcoupling}
\end{equation}
where, with $\varepsilon$ the $D^\ast$-meson polarisation vector,
\begin{equation}
\langle \pi(q) D(p) | D^*(p+q)\rangle =: g_{D^\ast \pi D} \; \varepsilon_{D^*}\! \cdot q
\end{equation}
It has been shown, however, that $\mathcal{O} (1/m_c)$ corrections are not negligible: applying Eq.\,\eqref{effcoupling} to both the strong decay $D^\ast D\pi$ and the (unphysical) process $B^\ast B\pi$ in the chiral limit \cite{ElBennich:2010ha, ElBennich:2012tp}, there is a material difference between the values of $\hat g$ extracted from either $g_{D^\ast \pi D}$ or $g_{B^\ast\pi B}$.  Another issue also arises: can $\hat g$ be realistically and unambiguously related to all the strong-interaction matrix elements mentioned above, $g_{D^\ast\pi D}$, $g_{D\rho D}$, etc., \emph{i.e}.\ is there a practically useful universal value of $\hat g$?

The study of interactions between charmed mesons and nuclear matter are a major piece of the proposed activities of the $\overline{\textrm P}$ANDA Collaboration at the future Facility for Antiproton and Ion Research (FAIR)  \cite{Wiedner:2011mf}; and it could also conceivably be pursued at the upgraded Jefferson Laboratory.  For example, low-momentum charmonia, such as $J/\psi$ and $\psi$, as well as $D^{(\ast )}$ mesons, can be produced by annihilation of antiprotons on nuclei \cite{Haidenbauer:2014rva} and in electroproduction from nuclei owing to enhancements of hadronic interactions at threshold \cite{Brodsky:2012zzb, E12-12-001, E12-12-006, Brodsky:2015zxu}.  Since charmonia do not share valence-quarks in common with the surrounding nuclear medium, competing mechanisms have been proposed to describe the influence of that medium on their propagation, \emph{e.g}.\ QCD van der Waals forces, arising from the exchange of two or more gluons between color-singlet states \cite{Peskin:1979va, Brodsky:1989jd}; and intermediate charmed hadron states \cite{Brodsky:1997gh, Ko:2000jx}, such that $\bar D^{(\ast)} D^{(\ast)}$ hadronic vacuum polarization components of the $J/\psi$ interact with the medium via meson exchanges \cite{Krein:2010vp}.

Model Lagrangians have also been employed to study interactions between open-charm mesons and nuclei in attempts to explore the possibility of charmed nuclear bound states \cite{Sibirtsev:1999js, Mizutani:2006vq, Haidenbauer:2007jq, Molina:2008nh, Haidenbauer:2010ch, Yamaguchi:2011xb}.  They are formulated with couplings between $D^{(\ast)}$ mesons and light pseudoscalar and vector mesons, which are typically derived via an $SU(4)$ extension of light-flavor chirally-symmetric Lagrangians.  Exotic states formed from a heavy meson and a nucleon were also investigated using heavy-meson chiral perturbation theory \cite{Yamaguchi:2011xb}.

One might also seek a direct computation of the in-medium properties of systems containing charm by using a bound-state formalism.  For example, the variation of the pion's mass, decay constant and elastic form factor as a function of nuclear density was recently studied in a light-front approach in Ref.\,\cite{deMelo:2014gea}.  In the longer term, however, one hopes to achieve a symmetry-preserving description of the in-medium properties of charm bound-states via a direct study of the relevant gap and Bethe-Salpeter equations subject to the inclusion of a chemical potential, as completed for the $\pi$- and $\rho$-mesons with simplified interaction kernels \cite{Maris:1997eg, Roberts:2000aa}.

In the context of this widespread interest in systems containing charm, herein we determine the general structure of the coupling of a $\rho$-meson, which is typically off-shell, to on-shell $D$- and $D^\ast$-mesons and compute the Poincar\'e-invariant amplitudes that arise, using a framework in which all elements are constrained by Dyson-Schwinger equation (DSE) studies in QCD.  This approach provides a simultaneous description of light- and heavy-quarks and the bound states they constitute as well as of their electromagnetic and heavy-to-light transition form factors and strong couplings in the so-called triangle digram approximation~\cite{ElBennich:2010ha,ElBennich:2012tp,Roberts:2000aa, Roberts:2007ji, Chang:2011vu, Bashir:2012fs, Cloet:2013jya, Horn:2016rip,ElBennich:2011py}.

The simplest case, \emph{viz}.\ the $D\rho D$ coupling, $g_{D\rho D}$, has previously been considered, with an estimate of the $SU(4)$ flavor-breaking pattern which relates $g_{D\rho D}$ to the well-constrained benchmark $\pi\rho \pi$ coupling and to the $K\rho K$ coupling, namely the ratios of couplings $g_{K\rho K}/g_{D\rho D}$ and $g_{K\rho K}/g_{\pi\rho \pi}$. In the case of exact $SU(4)$ symmetry, these ratios are respectively, $1$ and $1/2$, though it was found that $SU(4)$ flavor symmetry is broken at the 300-400\% level~\cite{ElBennich:2011py}.  Furthermore, in comparison with common monopole parametrizations \cite{Haidenbauer:2010ch}, the $D\rho D$ form factors computed in Ref.\,\cite{ElBennich:2011py} possess a larger zero-momentum coupling and are considerably softer.  These qualities could have a significant impact, \emph{e.g}.\ on predictions for $X(3872)$ production in heavy-ion collisions, which have sometimes used momentum-independent couplings based on $SU(4)$ symmetry \cite{Torres:2014fxa}.

Herein, therefore, we revisit and extend the study in Ref.\,\cite{ElBennich:2011py}, and compute the couplings $D \rho D$, $D^\ast\! \rho D$ and $D^\ast\! \rho D^\ast$.
%
We analyse the complete structure of the amplitudes and expose differences in the magnitude and functional forms of the five associated momentum-dependent couplings.

\section{Charm Amplitudes}
The couplings used in phenomenological Lagrangians can be related to amplitudes describing transitions between on-shell pseudoscalar and/or vector $D= c\bar f$ ($f=u,d$) mesons via emission of a $\rho$-meson, which is typically off-shell in practice. The transition amplitudes may be expressed in terms of incoming and outgoing momenta, $p_1$ and $p_2$ respectively, with $q=p_2-p_1$ being the $\rho$-meson momentum, and the helicity, $\lambda_M$, $M=\rho, D, D^\ast$, of the participating vector-mesons.  They can be used to define the dimensionless couplings $g_{D \rho D}(q^2)$ and  $g_{D^*\! \rho D}(q^2)$:
\begin{align}
g_{D \rho D}(q^2) & \, \bm{\epsilon}^{\lambda_\rho}\! \cdot\, p_1 :=  \langle  D(p_2) | \, \rho (q,\lambda_\rho)\,  | D (p_1 ) \rangle   \, ,  \label{eq1} \\
\nonumber
g_{D^*\!\rho D}(q^2)& \, \frac{1}{m_{D^*}}  \varepsilon^{\alpha\beta\mu\nu}\,
 \bm{\epsilon}_\alpha^{\lambda_{D^*}} \bm{\epsilon}_\beta^{\lambda_\rho} \ p_{1\mu}\, p_{2\nu} := \\
& \quad \langle  D^*(p_2,\lambda_{D^*}) | \, \rho (q,\lambda_\rho)\,  | D (p_1 ) \rangle\,.   \label{eq2}
\end{align}

Owing to the presence of three vector-mesons, two on-shell, the $D^*\!\rho D^*$ amplitude is more complicated, involving three couplings \cite{Arnold:1979cg, Hawes:1998bz, Bhagwat:2006pu}:
\begin{align}
\nonumber
& \langle  D^*(p_2,\lambda_{D^*}) | \, \rho (q,\lambda_\rho)\,  | D^* (p_1 ,\lambda_{D^*}) \rangle\\
& =   - \sum_{i=1}^{3} T_{\mu\rho\sigma}^{\,i} (p, q) \; g^i_{D^*\!\rho D^*}(q^2) \,
  \bm{\epsilon}_\mu^{\lambda_{\rho} } \bm{\epsilon}_\rho^{\lambda_{D^*}} \bm{\epsilon}_\sigma^{\lambda_{D^*} } ,
\label{DstarrhoDstar}
\end{align}
with
\begin{subequations}
\label{eq13}
\begin{eqnarray}
  T_{\mu\rho\sigma}^{\,1}(p,q) & = & 2\, p_\mu \, {\cal P}_{\rho\gamma}^T(p_1)\, {\cal P}_{\gamma\sigma}^T(p_2) \;,
 \label{Eq:HPT1}  \\
\nonumber
  T_{\mu\rho\sigma}^{\,2}(p,q) & = & - \left[q_\rho - p_{1 \rho} \, \frac{q^2}{2\,  m_{D^*}^2}\right]  {\cal P}_{\mu\sigma}^T(p_2) \\
&& \quad
    + \left[q_\sigma + p_{2\sigma} \, \frac{q^2}{2\, m_{D^*}^2}\right]  {\cal P}_{\mu\rho}^T(p_1) \; ,
 \label{Eq:HPT2}  \\
\nonumber
 T_{\mu\rho\sigma}^{\, 3}(p,q) &=& \frac{p_\mu}{m_{D^*}^2} \, \left[q_\rho - p_{1 \rho}\, \frac{q^2}{2\,  m_{D^*}^2}\right]  \\
 &&   \quad \times  \left[q_\sigma + p_{2 \sigma}\, \frac{q^2}{2\, m_{D^*}^2}\right] \ ,
\end{eqnarray}
\end{subequations}
where the four-momentum $p$ is defined by $p_1 =p- \tfrac{1}{2} q$ and $p_2 =p +  \tfrac{1}{2} q$, $p_1^2 =p_2^2 = -m_{D^{(*)}}^2$, ${\cal P}_{\gamma\sigma}^T$ is the standard transverse projection operator, and we have deliberately included a ``$-$'' sign in the definition of $T_{\mu\rho\sigma}^{\,2}$ so that $\{g^i_{D^*\!\rho D^*}(q^2) \geq 0, i=1,2,3\}$.
The amplitude in Eq.\,\eqref{DstarrhoDstar} can be written:
\begin{equation}
\langle  D^*(p_2) | \, \rho (q)\,  | D^* (p_1) \rangle =
\Lambda_{\mu\rho\sigma}(p,q)\, \bm{\epsilon}_\mu^{\lambda_{\rho} }\bm{\epsilon}_\rho^{\lambda_{D^*}}  \bm{\epsilon}_\sigma^{\lambda_{D^*}},
\end{equation}
in which case:
\begin{subequations}
\begin{eqnarray}
 p_{2\rho}\, \Lambda_{\mu\rho\sigma}(p,q) & = &  0  \, ,
\\
 p_{1\sigma}\, \Lambda_{\mu\rho\sigma}(p, q) & = &  0 \, ,
\\
 q_\mu\, \Lambda_{\mu\rho\sigma}(p,q) & = &   0 \, .
\end{eqnarray}
\end{subequations}
The tensor decomposition in Eq.\,\eqref{DstarrhoDstar} is not unique; more general structures have been proposed, with up to 14 form factors \cite{Bracco:2007sg}; but many of those are necessarily equal owing to Boson and charge conjugation symmetries.

The physical decay $\rho\to \pi\pi$ is also described by a matrix element like that in Eq.\,\eqref{eq1}.  In the present case, however, there are plainly no associated physical processes.  Notwithstanding that, a coupling of this sort is employed in defining $\rho$-mediated exchange interactions between a nucleon and kaons or $D$- mesons \cite{Haidenbauer:2007jq, Haidenbauer:2010ch}.  In such applications the off-shell $\rho$-meson's momentum is necessarily spacelike, and couplings and form factors may be defined once one settles on a definition of the off-shell $\rho$-meson.

We choose to employ a symmetry-preserving approach based upon DSE studies in QCD that provide sound results for mesons involving a heavy-quark \cite{Ivanov:1997yg, Ivanov:1998ms, Ivanov:2007cw, ElBennich:2009vx, ElBennich:2010ha, ElBennich:2012tp, Rojas:2014aka}.  In this approach, quark propagation is described by fully-dressed Schwinger functions, whose analytic structure is sufficient to ensure confinement \cite{Krein:1990sf}.  The dressing has a particularly significant effect on the properties of light-quarks, which are characterised by a strongly momentum-dependent running-mass whose impact on observables cannot adequately be captured by using a single constituent-like mass-value \cite{ElBennich:2008qa, ElBennich:2008xy, Chang:2009ae, Bashir:2012fs, ElBennich:2012ij}.

At leading-order in a systematic, symmetry-preserving truncation scheme \cite{Binosi:2016rxz}, one may express Eq.\,(\ref{eq1}) as
\begin{eqnarray}
\lefteqn{\hspace*{-6mm}  g_{D\rho D} \  \bm{\epsilon}_{\lambda_\rho}\!\cdot p_1  =    \tr \! \int\! \frac{d^4k}{(2\pi)^4} \, \Gamma_D (k; p_1)  S_c(k_c) }  \nonumber   \\
  &  \times &  \bar  \Gamma_D(k;-p_2) S_{l}(k_f')\, \bm{\epsilon}_{\lambda_\rho}^*\!\cdot \bar \Gamma_\rho(k; -q) S_{l}(k_f)  \,  ,
\label{mainamplitude}
\end{eqnarray}
where the trace is taken over color and Dirac indices;
$S_f(k_f)$ represents a dressed-quark propagator for the indicated flavor [we work in the isospin symmetric limit, so $l=u=d$];
$\Gamma_D$, $\Gamma_\rho$ are, respectively, Bethe-Salpeter amplitudes for the $D$- and $\rho$-mesons;
and momentum conservation requires $k_c = k+w_1 p_1,  k_f' =k+w_1 p_1 - p_2$,  $k_f =k-w_2p_1$, $f=u,d$, with the relative-momentum partitioning parameters satisfy $w_1 + w_2 = 1$.
The integral expressions for Eqs.\,\eqref{eq2} and \eqref{DstarrhoDstar} are obtained by changing $\Gamma_D (k;p)\, \to\, \bm{\epsilon}_{\lambda}\! \cdot \Gamma_{D^*} (k;p)$.  This impulse-like approximation has enjoyed widespread success, including applications such as that herein \cite{Ivanov:1998ms, Ivanov:2007cw, ElBennich:2010ha, ElBennich:2011py}.

\section{DSE framework}
The amplitude in Eq.\,\eqref{mainamplitude} can be computed once the dressed-quark propagators and meson Bethe-Salpeter amplitudes are specified.

\subsection{Dressed quark propagators}
\label{dressedquark}
The dressing of the quark (or antiquark) within a given meson is described by a gap equation, the solution of which has the general form
\begin{equation}
\label{SpAB}
S(p) = -i \gamma\cdot p\, \sigma_V(p^2) + \sigma_S(p^2)= Z(p^2) /[i\gamma\cdot p\,  + M(p^2)] \, .
\end{equation}
For light-quarks, it is a longstanding prediction that both the wave-function renormalization, $Z(p^2)$, and dressed-quark mass-function, $M(p^2)=\sigma_S(p^2)/\sigma_V(p^2)$, receive strong momentum-dependent modifications at infrared momenta: $Z$ is suppressed and $M$ enhanced.  These features are characteristic of dynamical chiral symmetry breaking (DCSB) and, plausibly, of confinement.\footnote{Eqs.\,(\protect\ref{ssm}), (\protect\ref{svm}) represent the quark propagator $S(p)$ as an entire function, which entails the absence of a Lehmann representation and is a sufficient condition for confinement \protect\cite{Krein:1990sf, Roberts:2007ji}.}  The importance of this infrared dressing has long been emphasized, \emph{e.g}.\ it is intimately connected with the appearance of Nambu-Goldstone modes \cite{Horn:2016rip}.  The predicted behavior of $Z(p^2)$, $M(p^2)$ has been confirmed in numerical simulations of lattice-regularized QCD \cite{Bhagwat:2003vw, Bowman:2005vx, Bhagwat:2006tu, Roberts:2007ji}.

Whilst numerical solutions of the quark DSE are readily obtained, the utility of an algebraic form for $S(p)$, when calculations require the evaluation of numerous integrals, is self-evident.  
Such an algebraic propagator must not only provide an accurate parametrization of extant DSE solutions at spacelike momenta, it must reproduce two crucial features of DSE studies;
namely, violation of reflection positivity and agreement with perturbative QCD at ultraviolet momenta neglecting anomalous dimensions for simplicity.
An efficacious parametrization, exhibiting all the above features and used extensively~\cite{Ivanov:1998ms, Ivanov:2007cw, Segovia:2015hra, Segovia:2015ufa}, is expressed via
\begin{subequations}
\label{SpropForm}
\begin{eqnarray}
\nonumber \bar\sigma_S(x) & =&  2\,\bar m \,{\cal F}(2 (x+\bar m^2))\\
&&  + {\cal
F}(b_1 x) \,{\cal F}(b_3 x) \,
\left[b_0 + b_2 {\cal F}(\epsilon x)\right]\,,\label{ssm} \\
\label{svm} \bar\sigma_V(x) & = & \frac{1}{x+\bar m^2}\, \left[ 1 - {\cal F}(2 (x+\bar m^2))\right]\,,
\end{eqnarray}
with $x=p^2/\lambda^2$, $\bar m$ = $m/\lambda$, ${\cal F}(x)= [1-\exp(-x)]/x$,
$\bar\sigma_S(x) = \lambda\,\sigma_S(p^2)$ and $\bar\sigma_V(x) = \lambda^2\,\sigma_V(p^2)$.  
The parameters were fixed by requiring a least-squares fit to a wide range of light- and heavy-meson observables, and take the values \cite{Ivanov:1998ms}:
\begin{equation}
\label{tableA}
\begin{array}{llcccc}
f &   \bar m_f& b_0^f & b_1^f & b_2^f & b_3^f \\\hline
u=d &   0.00948 & 0.131 & 2.94 & 0.733 & 0.185 
\end{array} \, .
\end{equation}
\end{subequations}
At a scale $\lambda=0.566\,$GeV, the current-quark mass is $m_u=5.4\,$MeV and one obtains the following Euclidean constituent-quark mass, defined as
\begin{equation}
\hat M^{E} = \{\sqrt{s} \, |\, s+M^2(s)=0,s>0\} = 0.36\,\mbox{GeV}.
\end{equation}
[\emph{N.B}.\ $\epsilon=10^{-4}$ in Eq.\,(\protect\ref{ssm}) acts only to decouple the large- and intermediate-$p^2$ domains.]

Whereas the impact of DCSB on light-quark propagators is significant, the effect diminishes with increasing current-quark mass [see, \emph{e.g}.\ Fig.\,1 in Ref.\,\cite{Ivanov:1998ms}].  This can be explicated by considering the dimensionless and renormalization-group-invariant ratio $\varsigma_f:=\sigma_f/M^E_f$, where $\sigma_f$ is a constituent-quark $\sigma$-term: $\varsigma_f$ measures the effect of explicit chiral symmetry breaking on the dressed-quark mass-function compared with the sum of the effects of explicit and dynamical chiral symmetry breaking.  Calculation reveals \cite{Roberts:2007jh}: $\varsigma_u = 0.02$, $\varsigma_s = 0.23$, $\varsigma_c = 0.65$, $\varsigma_b = 0.8$.  Plainly, $\varsigma_f$ vanishes in the chiral limit and remains small for light quarks, since the magnitude of their constituent mass owes primarily to DCSB.  On the other hand, for heavy quarks, $\varsigma_f\to 1$ because explicit chiral symmetry breaking is the dominant source of their mass.  Notwithstanding this, confinement remains important for the heavy-quarks.  These considerations are balanced in the following simple parametrized form for the $c$-quark propagator:
\begin{equation}
\label{SQ}
  S_c (k) = \frac{-i \gamma\cdot k + \hat M_c}{\hat M_c^2} {\cal F}(k^2/\hat M_c^2)\,,
\end{equation}
which implements confinement but produces a mo\-men\-tum-independent $c$-quark mass-function; namely, $\sigma_S^c(k^2)/\sigma_V^c(k^2)=\hat M_c$.  We use $\hat M_c = 1.32\,{\rm GeV}$ \cite{Ivanov:1998ms}.

\subsection{Bethe-Salpeter amplitudes}
A meson is described by the amplitude obtained from a homogeneous Bethe-Salpeter equation.  In solving that equation the simultaneous solution of the gap equation is required.  Since we have already chosen to simplify the calculations by parametrizing $S(p)$, we follow Refs.\,\cite{ElBennich:2010ha, Ivanov:1998ms, Ivanov:2007cw, ElBennich:2009vx} and also employ that expedient with $\Gamma_{D^{(*)}\!,\, \rho}$.

Regarding the $\rho$, DSE studies of light-vector mesons \cite{Pichowsky:1999mu, Jarecke:2002xd} indicate that, in applications such as ours, one may effectively use
\begin{subequations}
\label{BSAs}
\begin{equation}
\label{GV}
 \Gamma^\mu_\rho (k;P) =  \left ( \gamma_\mu -P_\mu\, \frac{\gamma\cdot P}{P^2} \right )   \frac{\exp (-k^2/ \omega_\rho^2) }{\mathcal{N}_\rho}  \, ;
\end{equation}
namely, a function whose support is greatest in the infrared.\footnote{The correct ultraviolet behaviour of meson Bethe-Salpeter amplitudes is $\sim 1/k^2$, up to logarithmic corrections, \emph{e.g}.\ Refs.\,\cite{Maris:1997tm, Maris:1999nt}.  However, experience reveals that this ``tail'' has no material impact on analyses such as that herein.
}
Similarly, for the charm mesons we choose:
\begin{equation}
\label{GH}
   \Gamma_D (k;P) = i \gamma_5 \, \frac{\exp (-k^2/\omega_D^2) }{\mathcal{N}_D} \,;
\end{equation}
and
\begin{equation}
  \Gamma^\mu_{D^*} (k;P) =  \left ( \gamma_\mu + P_\mu\, \frac{\gamma\cdot P}{M_{D^*}^2} \right )   \frac{\exp (-k^2/ \omega_{D^*}^2) }{\mathcal{N}_{D^*}}  \,  .
\label{GDstar}
\end{equation}
\end{subequations}
These \emph{Ans\"atze} represent a practical simplification, whose form derives from the understanding of meson Bethe-Salpeter amplitudes accumulated
during the past two decades~\cite{Maris:2003vk} and introduce three vector-meson width parameters: $\omega_\rho$, $\omega_D$, $\omega_{D^*}$.
Since we don't assume heavy-quark symmetry, $\omega_D \neq \omega_{D^*}$.

The quantities $\mathcal{N}_\rho$,  $\mathcal{N}_D$  and $\mathcal{N}_{D^*}$ in Eqs.\,\eqref{BSAs} are canonical on-shell normalization constants.  They are defined such that, \emph{e.g}.\
\begin{eqnarray}
  2\, P_\mu & = & \left [ \frac{\partial}{\partial K_\mu} \Pi(P,K) \right ]_{K=P}^{P^2=-m^2_D} \ ,
  \label{norm1} \\
   \Pi(P,K) & = &   \tr \int\!  \frac{d^4k}{(2\pi)^4} \, \bar \Gamma_D(k;-P) S_c (k+w_1K) \nonumber \\
   &  & \times \   \Gamma_D(k;P) S_l (k-w_2K) \,,
  \label{norm2}
\end{eqnarray}
for the $D$-meson, with analogous expressions for the $\rho$ and $D^*$ \cite{Ivanov:1998ms}.  Using Bethe-Salpeter amplitudes normalized in this way,
the width parameters may be fixed by computing the mesons' leptonic decay constants, $f_M$:
\begin{subequations}
\label{leptonicdecay}
\begin{eqnarray}
\label{psdecay}
P_\mu f_{D}  &=& \tr \int\!  \frac{d^4k}{(2\pi)^4} \, \gamma_5 \gamma_\mu\, \chi_{D}(k;P)\,, \\
\label{vecdecay}
m_V f_V & = & \tfrac{1}{3}\tr \int\!  \frac{d^4k}{(2\pi)^4} \,  \gamma_\mu\, \chi_V^\mu(k;P) \, ,
\end{eqnarray}
\end{subequations}
where $V=\rho$, $D^\ast$, $\chi_M= S_{f_1}(k+w_1P) \Gamma_M(k;P) S_{f_2}(k-w_2P)$, with flavors $f_{1,2}$ chosen appropriately, and then requiring agreement with experiment or reliable theoretical predictions.

The DSE approach to the calculation of hadron observables is Poincar\'e covariant provided the full structure of  hadron bound-state amplitudes is retained \cite{Maris:1997tm}.  However, we restrict ourselves to a simple one-covariant model for the amplitudes, with the goal of simultaneously describing a wide range of phenomena; and with omission of the full structure of amplitudes comes the complication that our results can be sensitive to the definition of the relativistic relative momentum.  Every study that fails to retain the full structure of the Bethe-Salpeter amplitude shares this complication.  To proceed, we must therefore specify the relative momentum in Eq.\,\eqref{mainamplitude} and its analogues, and in Eqs.\,\eqref{norm2}--\eqref{leptonicdecay}.   When a heavy-quark line is involved, we allocate a fraction, $w_1$, of the heavy-light-meson's momentum to that heavy-quark; consequently, $w_2$ is the momentum fraction carried by the light quark.  A natural choice is
\begin{equation}
w_1^c = \frac{\hat M_c}{\hat M_c + \hat M_{l}} \quad \Rightarrow \quad w_1^c =0.78, \ w_2^c = 0.22 \, ,
\end{equation}
which allocates most, but not all, of the heavy-light-meson's momentum to the $c$-quark.  We stress that in a Poincar\'e invariant calculation, no physical observable can depend on the choice of momentum partitioning; but that feature is compromised in our approach and any sensitivity to the partitioning is an artifact that owes to our simplifications.  
(This issue is further elaborated elsewhere \cite{Gomez-Rocha:2015qga, Gomez-Rocha:2016cji})

\section{Numerical results}
Using Eq.\,\eqref{norm2} and its analogues, along with Eqs.\,\eqref{leptonicdecay} we obtain the results in Table~\ref{widths}.  Notably, using these values one obtains \cite{ElBennich:2012tp} $g_{D^\ast \pi D} = 18.7^{+2.5}_{-1.4}$, which may be compared with $g_{D^\ast \pi D }^\mathrm{exp.}=17.9\pm1.9$ \cite{Anastassov:2001cw}; $16.92\pm 0.13 \pm 0.14$~\cite{Lees:2013uxa}.  It is also worth remarking that the momentum-space widths determined in this way are consistent with intuition [all lengths measured in fm]:
\begin{equation}
\ell_D : =1/\omega_D = 0.183 < \ell_{D^*}  = 0.244 < \ell_\rho  = 0.352\,,
\end{equation}
{\em i.e.\/} by a rough measure, a vector meson is larger than  a pseudoscalar meson; and charm states are smaller than light-quark states.

\begin{table}[t]
\caption{Leptonic decay constants computed using $m_D=1.87\,$GeV,  $m_{D^*}=2.01\,$GeV, $m_\rho=0.77\,$GeV.  In connection with the $\rho$-meson, we use $w_2^\rho = 0.38$; and, experimentally, $f_\rho=0.216\,$GeV, obtained from the $e^+ e^-$ decay width \cite{Agashe:2014kda}.
 [All tabulated entries in GeV and $f_\pi = 0.131\,$GeV with this normalization.]
\label{widths}
}
\begin{center}\begin{tabular*}
{\hsize}
{
l@{\extracolsep{0ptplus1fil}}
l@{\extracolsep{0ptplus1fil}}
l@{\extracolsep{0ptplus1fil}}}\hline
$M$ & $\omega_M$ & $f_M$ \\\hline
$D$ & $1.08\pm 0.1$ & $0.206\pm 0.009$ \cite{Eisenstein:2008aa} \\
$D^\ast$ & $0.81\pm 0.15$ & $0.245\pm 0.020$ \cite{Becirevic:1998ua} \\
$\rho$ & $0.66$ & $0.22$ \\\hline
\end{tabular*}
\end{center}
\end{table}

\begin{figure}[t]
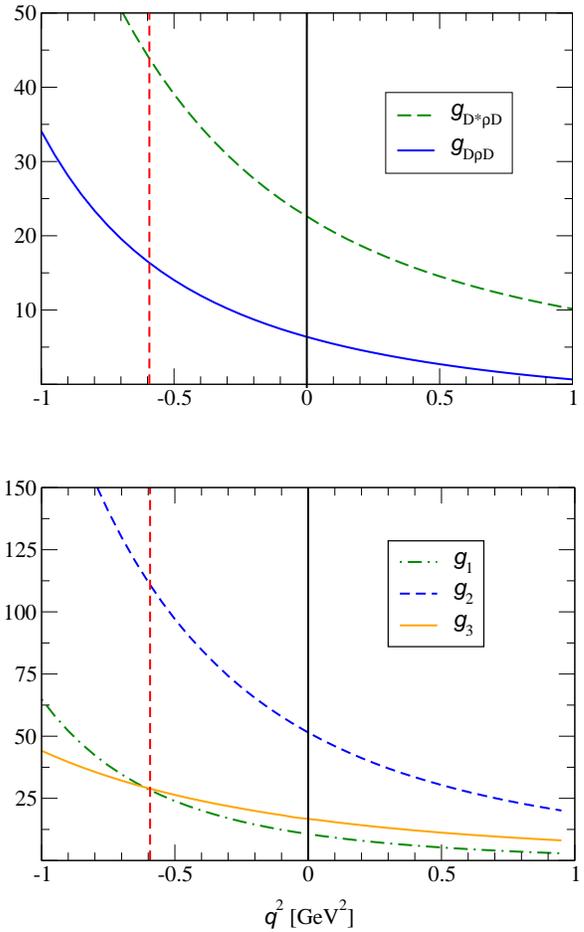

\includegraphics[width=0.42\textwidth]{F1A.eps}
\vspace*{3.3ex}

\hspace*{-0.7em}\includegraphics[width=0.43\textwidth]{F1B.eps}
\caption{(Color online)
\label{Fig1}
\emph{Upper panel}.
The dimensionless couplings $g_{D \rho D}$ (solid, blue) and $g_{D^*\! \rho D}$  (dashed, green), computed as a function of the $\rho$-meson's four-momentum squared, with the $D$ and $D^*$ mesons on-shell.    \emph{N.B}.\ $g_{D^*\! \rho D}$ rather than $g_{D^*\! \rho D}/m_{D^*}$ is plotted and $q^2 > 0$ is spacelike with our Euclidean metric.
\emph{Lower panel}.  The three vector-couplings in Eq.\,\eqref{DstarrhoDstar}, $\{g_{D^*\! \rho D^*}^i \equiv  g_i, i=1,2,3\}$.  Again, the $D^*$-mesons are on-shell.
In both panels the dashed (red) vertical line marks the $\rho$-meson on-shell point.}
\end{figure}

With the meson width-parameters now fixed, the couplings $g_{D \rho D}$,  $g_{D^*\! \rho D}$ and $\{g^i_{D^*\!\rho D^*},i=1,2,3\}$, Eqs.~\eqref{eq1}--\eqref{eq13} respectively, can now be computed using Eq.\,\eqref{mainamplitude} and its analogues.
The coupling $g_{D \rho D}$ was discussed elsewhere \cite{ElBennich:2011py} in connection with $SU(4)$ flavor-symmetry breaking; but we nevertheless recompute it herein, taking this opportunity to improve on the numerical integration method used previously.  Both dimensionless couplings $g_{D \rho D}$ and $g_{D^*\! \rho D}$ are plotted as a function of the off-shell $\rho$-meson momentum, $q^2$, in Fig.\,\ref{Fig1} (upper panel): they are smooth and monotonically decreasing as $q^2$ increases away from the on-shell point $q^2=-m_\rho^2=:-\hat s$.

On $s=q^2 \in [-\hat s,\hat s]$, our results are reliably interpolated by the following functions:
\begin{eqnarray}
 g_{D \rho D} (s)    & = &  \frac{6.37 -5.03 s}{1.0+ 0.80 s + 0.14 s^2} \\
 g_{D^*\!\rho D} (s)  & = &  \frac{22.60 + 0.35 s}{1.0+ 1.0 s + 0.28 s^2} \ . \label{interpol}
\end{eqnarray}
Notably, the coupling $g_{D^*\! \rho D}$ is, on average, four-times larger than $g_{D \rho D}$ on this domain, something which can primarily be attributed to differences in the $D$ and $D^*$ canonical normalization constants and hence, indirectly, to $f_{D^\ast} > f_D$ (see Table~\ref{widths}).  Moreover, 
\begin{equation}
g_{D \rho D} (0) \approx 6.4  < g_{D^*\! \pi^0 D} \approx  13 <  g_{D^*\! \rho D}(0) \approx 23\,.
\end{equation}

In principal, the coupling between three vector-states could generate a complicated tensor structure; but, as we noted in connection with Eq.\,\eqref{DstarrhoDstar}, symmetries reduce the number of independent $D^\ast \rho D^\ast$ couplings to just those three depicted in Fig.\,\ref{Fig1}, lower panel.  On the domain $s\in [-\hat s,\hat s ]$, these curves are reliably interpolated by the following functions:
\begin{subequations}
\begin{eqnarray}
g_1 (s)   & = &  \frac{10.52 - 2.0 s}{1.0 + 1.25 s + 0.44 s^2} \\
g_2 (s)   & = &  \frac{51.53 - 8.41s}{1.0 + 0.98 s + 0.27 s^2} \\
g_3 (s)   & = &  \frac{16.80 - 2.02s}{1.0 + 0.73 s + 0.15 s^2}  \ .
\end{eqnarray}
\end{subequations}
For clarity, we list all $q^2=0$ coupling values in Table~\ref{TAB1}.

All couplings are again smooth, monotonically decreasing functions of their argument; but there are notable quantitative differences between their magnitudes and damping rates.  For example, averaged on $s\in [-\hat s,\hat s]$,
\begin{equation}
\bar g_2(s) \approx 3 \, \bar g_3(s) \approx 5 \, \bar g_1(s)\,.
\end{equation}
Such relative strengths are of the same magnitude as those found in the $\rho$-meson elastic form factor \cite{Roberts:2011wy}.  Moreover,
\begin{equation}
\frac{g_1(-\hat s)}{g_1(\hat s)} \approx 6 \,,
\frac{g_2(-\hat s)}{g_2(\hat s)} \approx 4 \,,
\frac{g_2(-\hat s)}{g_3(\hat s)} \approx 3 \,.
\end{equation}
It is notable that $\bar g_3(s) \approx 0.7\,\bar g_{D^*\!\rho D} (s)$, \emph{i.e}.\ this one of the $D^\ast \rho D^\ast$ couplings is similar in magnitude to the $D^*\!\rho D$ coupling.
We find in addition that $g_3 (-\hat s) \approx g_1(-\hat s)$.  This approximate equality is not a consequence of symmetries; but it is fairly insensitive to a variation in $\omega_\rho$, the $\rho$-meson width parameter.

\begin{table}[t]
\caption{Dimensionless couplings computed at momentum transfer $q^2=0$, with the $D$ and $D^*$
mesons on-shell.   For additional comparison, we note that the same computational framework yields \cite{ElBennich:2010ha} $g_{D^\ast \pi D} = 18.7^{+2.5}_{-1.4}$.  (Using a neutral-pion normalisation, this corresponds to $g_{D^\ast \pi^0 D}=13.2^{+1.8}_{-1.0}$.) }
\label{TAB1}
\begin{center}
\begin{tabular*}
{\hsize}
{
c@{\extracolsep{0ptplus1fil}}
c@{\extracolsep{0ptplus1fil}}
c@{\extracolsep{0ptplus1fil}}
c@{\extracolsep{0ptplus1fil}}
c@{\extracolsep{0ptplus1fil}}}\hline
  $g_{D \rho D}$ & $g_{D^*\!\rho D}$ & $g_{D^*\!\rho D^*}^1$ & $g_{D^*\!\rho D^*}^2$  &  $g_{D^*\!\rho D^*}^3$ \\ \hline
    6.37 & 22.6  & 10.5  & 51.5  &  16.8 \\  \hline
\end{tabular*}
\end{center}
\end{table}

We close this section by remarking that errors on our computed couplings owe chiefly to uncertainties in the values of the weak-decay constants and the momentum partitioning parameters, both of which translate into 
uncertainties in the width parameters characterising our simplified meson Bethe-Salpeter amplitudes.\footnote{ 
Exemplifying, with a $\pm15\%$  variation in $w_1^c$: the leptonic decay constants, $f_D$ and $f_{D^*}$, change by less-than $3\%$; and, on the domain depicted, the form factors exhibit a mean 
relative shift of less-than $9\%$.  Thus, on a sizable neighborhood centered on the model-defining value of $w_1^c$, our predictions are qualitatively and semi-quantitatively unchanged.  For comparison, if one attributes 
{\em all} the $D$-meson's momentum to the heavy quark ($w_1^c=1$), none of the results change by more than 10\%.  At the other extreme, for  $w_1^c=0$, the form factors are moderately altered ($\sim 30$\%), 
but it is not possible to explain $f_D$ nor $g_{D^*\!D\pi}$.  } 
The analysis shares these characteristics with many other leading-order DSE computations and hence we judge that our predictions are accurate at the level of 
$\lesssim 15$\%, as is typical of analyses using this truncation~\cite{Xu:2015kta}.

\section{Comparisons with other evaluations}
Lattice-regularized QCD (lQCD) results on electromagnetic form factors of $D$- and $D^*$-mesons have been used as constraints on a vector dominance model in order to infer \cite{Can:2012tx} two of the $q^2=0$ couplings listed in Table~\ref{TAB1}:
\begin{equation}
\begin{array}{cc}
g_{D \rho D}^\mathrm{Latt.} & g_{D^*\! \rho D^*}^\mathrm{Latt.}\\
%
4.84(34) & 5.95(56)
\end{array}.
\label{latticeguesses}
\end{equation}
The $D \rho D$ value is similar to both our prediction and the result obtained using QCD sum rules $g_{D \rho D}^\mathrm{SR} = 2.9\pm 0.4$ \cite{Bracco:2001dj, Bracco:2011pg}, albeit closer to our result; but the $D^*\! \rho D^*$ strength is roughly 50\% of our computed value for the weakest of the three independent couplings, $g_{D^*\!\rho D^*}^1(0)$, which is associated with the tensor structure analysed in the lQCD study.

The $D^\ast \pi D$ coupling is also reported in Ref.\,\cite{Can:2012tx}:
\begin{equation}
\label{gDspiD}
g_{D^\ast \pi D}^\mathrm{Latt.} =16.23(1.71)\,,
\end{equation}
obtained from a matrix element involving an axial-vector current.  Within errors, this value agrees with the result listed in the caption of Table~\ref{TAB1}, which was obtained using the same framework as ours.  Converted to the neutral-pion normalisation, Eq.\,\eqref{gDspiD} corresponds to $g_{D^\ast \pi^0 D}^\mathrm{Latt.}=11.48(1.21)$, which is also similar to the sum rules result: $g_{D^\ast \pi D}^\mathrm{SR} = 9.9 \pm 1.0$ \cite{Bracco:2011pg}, although larger, as was the case with $D \rho D$ coupling.

Sum rules have also been used to estimate the $D^*\! \rho D$ coupling \cite{Rodrigues:2010ed, Bracco:2011pg}: $\tilde g_{D^*\! \rho D}^\mathrm{SR} :=g_{D^*\! \rho D}^\mathrm{SR} (-m_\rho^2)/m_{D^\ast}= 4.3 \pm 0.9\,$GeV$^{-1}$.  This value is factor of five less than we predict, \emph{viz}.\ using Eq.\,\eqref{interpol}, $g_{D^*\! \rho D} (-m_\rho^2) /m_{D^\ast} \approx 22\,$GeV$^{-1}$.

A similar sum rules analysis was employed to evaluate the $D^*\!\rho D^*$ couplings \cite{Bracco:2007sg, Bracco:2011pg}.  It concentrated on one particular tensor structure [$\delta_{\rho \sigma} q_\mu$, in the conventions of Eq.\,\eqref{DstarrhoDstar} herein] and, as with the analyses of other couplings, employed notions of quark-hadron duality.  Namely, the $D^*\!\rho D^*$ amplitude in Eq.\,\eqref{DstarrhoDstar} that is associated with the chosen tensor structure is computed twice: once using sum rules and again using a phenomenological Lagrangian \cite{Lin:1999ad}.  The resulting expressions are equated; and then a double Borel transformation performed on the variables $p_1^2$, $p_2^2$, which express on-shell quark-mass singularities in the triangle-diagram sum-rules computation and $D^\ast$-meson mass-poles on the phenomenological Lagrangian side.  It is argued \cite{Bracco:2007sg, Bracco:2011pg} that this Borel step improves matching: plainly, it works to mask the mismatch between the location of singularities in the two equated expressions.
The procedure yields a single coupling at spacelike momenta, $q^2>0$, which is extrapolated into the timelike region using various parametrizations \cite{Bracco:2011pg}, resulting in a value $g_{D^*\! \rho D^*}^\mathrm{SR}(-m_\rho^2)=4.7 \pm 0.2$.  
%
This value is a factor of six smaller than the weakest of the three independent couplings that we have computed: $g^1_{D^*\! \rho D^*}(-m_\rho^2) \approx 28$.
It is worth highlighting in this connection that whereas we predict:
\begin{equation}
g^1_{D^*\! \rho D^*}(-m_\rho^2) \sim 2\, g_{D^\ast \pi^0 D} \sim 4\, g^{\rm exp}_{\rho\pi\pi}\,,
\end{equation}
where the on-shell $\rho\to \pi\pi$ coupling $g^{\rm exp}_{\rho\pi\pi}(-m_\rho^2) = 6.0$ \cite{Agashe:2014kda}, the sum rules analyses produce:
\begin{equation}
g^\mathrm{SR}_{D^*\! \rho D^*}(-m_\rho^2) \sim \tfrac{1}{2}\, g^\mathrm{SR}_{D^\ast \pi^0 D}  \sim g^{\rm exp}_{\rho\pi\pi}\,.
\end{equation}

Evidently, the sum rules results reviewed herein are uniformly and significantly smaller than our predictions.  Underlying the sum rules analyses is the assumption of free-particle structure for the quark propagators appearing in the triangle diagram, which entails the presence of (unphysical) quark production thresholds but enables the use of dispersion relations to express the operator-product-expansion side of the equation.  In addition, the meson$\leftrightarrow$quark$+$antiquark vertices are bare.
In contrast, we employ dressed-quark propagators, Eqs.\,\eqref{SpAB}, \eqref{SpropForm}, which ensure that the quarks appearing in the triangle diagram are confined, \emph{i.e}.\ the propagators do not possess free-particle poles; and we use Bethe-Salpeter amplitudes, Eqs.\,\eqref{BSAs}, to express the meson$\leftrightarrow$quark$+$antiquark correlations.  It is probable that these fundamental dissimilarities in the basic assumptions are largely responsible for the marked discrepancies between our results and those collected in Ref.\,\cite{Bracco:2011pg}.
Nevertheless, a deeper exploration of the differences is merited, with an aim of reconciling them.

\section{Conclusion}
In the heavy-quark limit there is a single universal quantity $\hat g$ that describes the coupling between heavy-light mesons and their purely light-quark counterparts; and it is plausible that this quantity can be extracted with some accuracy from theoretical analyses of the $ g_{B^*\!\pi B}$ coupling.  However, our study indicates that $\hat g$ does not play a practically useful role in the description of such interactions when they involve $D$- and $D^\ast$-mesons.  Indeed, couplings between $D$-, $D^\ast$-mesons and $\pi$-, $\rho$-mesons can differ by almost an order-of-magnitude, and they also exhibit different evolution with light-meson virtuality [Table~\ref{TAB1} and Fig.\,\ref{Fig1}].

It seems worthwhile to explore the impact of our findings on a wide variety of predictions for charmed meson observables, such as those described in the Introduction that are based on phenomenological meson-exchange models.

\begin{acknowledgments}
We are grateful for valuable communications and discussions with M.\,A.~Ivanov, F.~Navarra, M.~Nielsen and G.~Krein.
B.~El-Bennich acknowledges support from CNPq fellowship no.\,301190/2014-3.
This work was also supported by
FAPESP grant no.\,2013/16088-4
U.S.\ Department of Energy, Office of Science, Office of Nuclear Physics, under contract no.\,DE-AC02-06CH11357;
Patrimonio Aut\'{o}nomo Fondo Nacional de Financiamiento para la Ciencia, la Tecnolog\'{\i}a y la Innovac\'{\i}on, Francisco Jos\'e de Caldas; and Sostenibilidad-UDEA 2014–2015.
\end{acknowledgments}


\end{document}